\documentclass{mn2e}
\usepackage{epsfig}

\newif\ifAMStwofonts

\newcommand{\target}{XTE~J1859+226}
\newcommand{\HST} {\textit{HST}}
\newcommand{\XTE} {\textit{RXTE}}
\newcommand{\RXTE}{\textit{RXTE}}

\title[The Evolving Accretion Disc in XTE J1859+226] 
{The Evolving Accretion Disc in the Black Hole X-ray
Transient XTE J1859+226\thanks{Based on
observations made with the NASA/ESA Hubble Space Telescope, obtained
at the Space Telescope Science Institute, which is
operated by the Association of Universities for Research in Astronomy, 
Inc., under NASA contract NAS 5-26555. These observations are associated with
proposal GO\,8245.}}

\author[R. I. Hynes et al.]
       {R. I. Hynes$^1$\thanks{e-mail: rih@astro.soton.ac.uk}, 
	C. A. Haswell$^2$,
        S. Chaty$^2$, 
	C. R. Shrader$^3$,
	W. Cui$^{4}$\\
$^1$Department of Physics and Astronomy, University of Southampton, 
    Southampton, SO17 1BJ\\
$^2$Department of Physics and Astronomy, The Open University, Walton
    Hall, Milton Keynes, MK7 6AA\\
$^3$Laboratory for High-Energy Astrophysics, NASA
    Goddard Space Flight Center, Greenbelt, MD 20771, USA\\
$^4$Department of Physics, Purdue University, 1396 Physics Building,
    West Lafayette, IN 47907-1396, USA\\
} 
\date{Accepted 2001 November 16.
      Received 2001 November 14;
      in original form 2001 September 10}

\pagerange{\pageref{firstpage}--\pageref{lastpage}}
\pubyear{2001}

\begin{document}
\maketitle
%
%
\begin{abstract}
We present \HST, \RXTE, and UKIRT observations of the broad band
spectra of the black hole X-ray transient \target\ during the decline
from its 1999--2000 outburst.  Our UV spectra define the 2175\,\AA\
interstellar absorption feature very well and based on its strength we
estimate $E(B-V)=0.58\pm0.12$.  Hence we deredden our spectra and
follow the evolution of the spectral energy distribution on the
decline from outburst.  We find that the UV and optical data, and the
X-ray thermal component when detectable, can be fit with a simple
blackbody model of an accretion disc heated by internal viscosity and
X-ray irradiation, and extending to close to the last stable orbit
around the black hole, although the actual inner radius cannot be well
constrained.  During the decline we see the disc apparently evolving
from a model with the edge dominated by irradiative heating towards
one where viscous heating is dominant everywhere.  The outer disc
radius also appears to decrease during the decline; we interpret this
as evidence of a cooling wave moving inwards and discuss its
implications for the disc instability model.  Based on the
normalisation of our spectral fits we estimate a likely distance range
of 4.6--8.0\,kpc, although a value outside of this range cannot
securely be ruled out.
\end{abstract}
%
%
%
%
\section{Introduction}
\label{IntroSection}
Black hole X-ray transients (BHXRTs), also referred to as X-ray novae
and soft X-ray transients, are low-mass X-ray binaries in which long
periods of quiescence, typically decades, are punctuated by very
dramatic X-ray and optical outbursts, often accompanied by radio
activity (Tanaka \& Shibazaki 1996; Cherepashchuk 2000).  In a typical
outburst, the X-ray emission is dominated by thermal emission from the
hot inner accretion disk, and UV, optical, and IR emission is thought
to be produced by reprocessing of X-rays, predominantly by the outer
disc.  Studies of the UV--IR can thus tell us about the structure of
the outer disc and the effect of irradiation upon it.

\target\ was discovered on 1999 October 9 by the \XTE/ASM (Wood et
al.\ 1999).  A 15th magnitude optical counterpart was identified by
Garnavich, Stanek \& Berlind (1999).  Wagner et al.\ (1999) provided
spectroscopic confirmation finding a spectrum typical of BHXRTs in
outburst.  The source was also detected in the radio (Pooley \&
Hjellming 1999) and $\gamma$-ray (McCollough \& Wilson 1999; dal Fiume
et al.\ 1999) bands.  Subsequent observations of the $R\sim23$
quiescent optical counterpart have confirmed an orbital period of
$\sim9.1$\,hrs (Garnavich \& Quinn 2000; Sanchez-Fernandez et al.\
2000; Filippenko \& Chornock 2001) and a radial velocity analysis has
yielded an exceptional mass function of
$f(M)=(7.4\pm1.1)$\,M$_{\odot}$ (Filippenko \& Chornock 2001), the
largest of all the BHXRTs.  This makes it one of the most securely
identified Galactic black holes.

\begin{figure}
\begin{center}
\epsfig{angle=90,width=3.4in,file=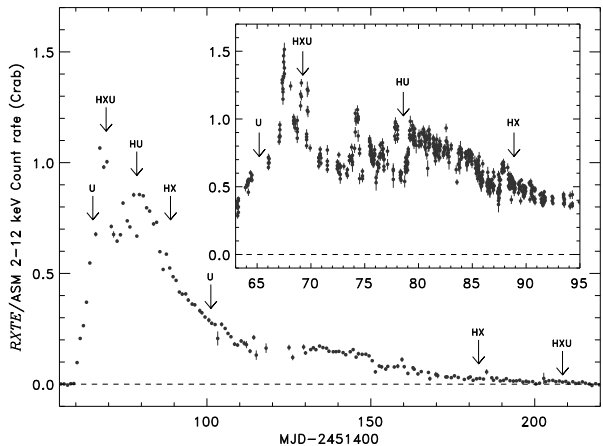}
\caption{\XTE/ASM lightcurve of \target\ in outburst based on
quick-look results provided by the ASM/\XTE\ team.  The main figure
shows one-day averages of the whole period we cover; the inset shows
individual 90\,s dwells from the early flaring period.  Times of
observations are marked with arrows, annotated H (\HST), X (\RXTE)
and/or U (UKIRT).  
}
\label{ASMFig}
\end{center}
\end{figure}

During the outburst a series of coordinated \HST, \RXTE\ and UKIRT
observations were made.  The times of the observations are marked
above the \RXTE/ASM lightcurve in Fig.~\ref{ASMFig}.  Some of the
results on X-ray timing have already been presented by Cui et al.\
(2000).  A more thorough analysis of the outburst lightcurves and
variability will be presented by Ioannou et al.\ (in preparation).  A
further work will discuss the emission line spectrum and detailed
X-ray spectral modelling.  Here we focus on the evolution of the
ultraviolet-optical-infrared (UVOIR) spectral energy distribution
(SED) and the implied evolution of the accretion disc.
%
%
\section{Observations}
\label{ObsSection}
\subsection{\HST\ Observations}
\HST\ observations of \target\ were made using the Space Telescope
Imaging Spectrograph (STIS; Leitherer et al.\ 2001) in five visits
spanning 1999 October 18 to 2000 March 5.  Unfortunately no
observations were obtained between 1999 November 5 and 2000 February 8
due to an \HST\ gyro failure and the target's proximity to the Sun.
At each epoch we obtained full but not simultaneous wavelength
coverage from 1150\,\AA\ to 10200\,\AA\ using the G140L and G230L MAMA
modes and the G430L and G750L CCD modes.  A large aperture
($52\times0.5$\,arcsec) was used to ensure photometric accuracy.  Full
details are given in Table \ref{ObsTable}.

\begin{table}
\caption{Log of \HST/STIS observations of \target.}
\label{ObsTable}
\begin{center}
\begin{tabular}{lcrl}
\hline
Date & Start     & Exp.\    & Grating\\
     & time (UT) & time (s) &        \\
\noalign{\smallskip}
1999 Oct 18 & 02:19:14 & 2250 & G140L \\
            & 03:41:40 & 2300 & G230L \\
            & 05:18:42 & 2300 & G140L \\
            & 06:05:57 &  250 & G230L \\
            & 06:55:22 &  300 & G430L \\
            & 07:06:47 &  200 & G750L \\
            & 08:32:06 & 2300 & G140L \\
\noalign{\smallskip}			
1999 Oct 27 & 19:55:24 & 2250 & G140L \\
            & 21:20:00 & 1350 & G140L \\
            & 21:52:16 &  300 & G430L \\
            & 22:03:41 &  203 & G750L \\
            & 22:57:02 & 2300 & G230L \\
\noalign{\smallskip}			
1999 Nov 6  & 19:57:20 & 2200 & G230L \\
            & 21:21:59 &  850 & G140L \\
            & 21:44:17 &  400 & G430L \\
            & 21:57:22 &  659 & G750L \\
\noalign{\smallskip}			
2000 Feb 8  & 17:19:52 & 1500 & G140L \\
            & 18:31:16 & 1150 & G140L \\
            & 19:00:12 &  450 & G430L \\
            & 19:14:07 &  253 & G750L \\
            & 20:08:38 & 2800 & G140L \\
            & 21:44:57 & 2800 & G230L \\
            & 23:21:55 & 2800 & G140L \\
\noalign{\smallskip}			
2000 Mar 5  & 06:39:52 & 1500 & G140L \\
            & 07:53:57 &  960 & G140L \\
            & 08:19:43 &  450 & G430L \\
            & 08:33:38 &  250 & G750L \\
            & 09:31:15 & 2760 & G140L \\
            & 11:07:32 & 2760 & G230L \\
            & 12:44:26 & 2760 & G140L \\
\hline
\end{tabular}
\end{center}
\end{table}

To examine the target's SED we used one-dimensional spectra from the
standard pipeline data products.  The photometric calibration accuracy
is estimated at 4\,percent for MAMA modes and 5\,percent for CCD modes
(Leitherer et al.\ 2001).  There is a further uncertainty in matching
different spectra due to source variability between exposures.

We constructed a continuum SED by masking out all significant emission
and absorption features (interstellar features were identified with
the aid of Blades et al.\ 1988) and then binning to
$\Delta\log\nu=0.01$ for the first three visits and
$\Delta\log\nu=0.02$ for the two later ones.  As the pipeline did not
remove all cosmic rays from CCD data, we took the median of all
continuum points within the bin for these data.  For MAMA data a
straight average was taken.  A further complication is that G750L data
suffer from significant fringing above $\sim8300$\,\AA, even though
contemporaneous flat fields were taken.  This is averaged out within
the bins, however.
\subsection{\RXTE\ Observations}

\begin{table*}
\caption{Results from UKIRT/UFTI observations of \target.}
\label{IRTable}
\begin{center}
\begin{tabular}{lcccc}
\hline
Date & JD & $J$ & $H$ & $K$ \\
\noalign{\smallskip}
1999 Oct 14 & 2451465.69 & $14.508\pm0.011$ & $14.231\pm0.014$ & $13.833\pm0.013$ \\
1999 Oct 18 & 2451469.69 & --               & --               & $14.516\pm0.012$ \\
1999 Oct 27 & 2451478.69 & $15.080\pm0.015$ & $14.689\pm0.010$ & $14.469\pm0.023$ \\
1999 Nov 19 & 2451501.69 & $15.504\pm0.008$ & $15.232\pm0.009$ & $15.033\pm0.025$ \\
2000 Mar 5  & 2451609.17 & $17.240\pm0.029$ & $16.935\pm0.027$ & $16.673\pm0.029$ \\
\noalign{\smallskip}
\hline
\end{tabular}
\end{center}
\end{table*}

We observed \target\ with the \RXTE\ proportional Counter Array (PCA)
and High-Energy Timing Experiment (HEXTE) at various epochs selected
to coincide with the \HST\ visits. Results presented here are from two
specific epochs at which well-characterised thermal X-ray components
are seen; 1999 October 18 (observation ID 40122-01-01-03) and 1999
November 6 (observation ID 40122-01-03-01). We have not included
analysis of the HEXTE data or later visits here, since these do not
contribute to our modelling of the accretion disc emission.  The
source intensity was about 2400 and 800 cts/sec per PCU at the
respective epochs, and the exposure times were 3100 and 1600\,s. We
used the standard-2 data (128 spectral channels, 16\,s
accumulations), selecting sub-intervals when the number of detectors
on remained constant (about 90\,percent of the total time) to form
128-channel detector count spectra. A subset of these channels,
corresponding to about 3--20\,keV were used in subsequent model
fitting. Background rates were estimated using the epoch-4 models, and
response matrices were generated using the current calibration files
and response-matrix generation software, all from the {\sc HEAsoft}
5.1 release.

The $\sim3-20$\,keV SEDs are suggestive of a high-soft state (i.e.\
high $\dot{m}$) accreting black hole, in that they consist of a
thermal disc component with a characteristic energy of about 1\,keV,
superimposed on a power law, probably formed by Comptonisation, with
index $\Gamma \simeq 2.5$, extending to higher energies.  However, the
ratio of the power law to disc components appears to be relatively
high for the first observation.  The Galactic hydrogen column density
is not well constrained by these data, and it was assumed to be in the
range of about $0.3-0.8 \times 10^{22}$\,cm$^{-2}$ (Markwardt et al.\
1999; dal Fiume et al.\ 1999), i.e.\ it was treated as a variable
parameter of our fitting, but constrained to that approximate range.
Fitting the data to a disk-blackbody model plus a power law leads to
values of $kT_{\rm in} \simeq 1.1$ and $0.9$\,keV and $\Gamma \simeq
2.5$ and $1.9$ at the two successive epochs. These hardening of the
power-law component is suggestive of the onset of a transition towards
the low-hard state. Indeed, some spectra from the later epochs of our
campaign are consistent with a low-hard-state source, with no
discernible disc contribution.
\subsection{UKIRT Observations}
Near-infrared service observations of \target\ were carried out at the
United Kingdom Infrared Telescope (UKIRT) 3.8\,m telescope using the
$1024\times1024$ pixel UFTI (1--2.5\,$\mu$m) camera (pixel scale
0.09\,arcsec) between 1999 October 14 (Chaty et al.\ 1999) and 2000
March 5.  The broad band filters $J$ (1.17--1.33\,$\mu$m), $H$
(1.49--1.78\,$\mu$m) and $K$ (2.03--2.37\,$\mu$m) were used.  The
conditions were photometric for most of the observations, the seeing
being typically 0.8\,arcsec.  A full log of the results is given in
Table~\ref{IRTable}.

Each exposure of the object is the average of 1\,min integration time
frames, repeated 9 times by offseting the images by 1\,arcmin to the
North-West, North-East, South-East and South-West from the central
position.  The final image is constructed by co-adding and median
filtering those individual frames.  Total exposure times were
therefore 9\,min in each filter.  The images were processed using {\sc
iraf} reduction software. Each of the images were corrected by a
normalized flat field, and sky-subtracted using a sky image created by
combining with a median filter a total of 9 consecutive images.  The
data were then analysed using the {\sc iraf} reduction task {\sc
apphot}, taking different apertures depending on the photometric
conditions.

Absolute photometry was performed using a nearby NICMOS photometric
standard star from the new system of faint near-infrared standard
stars (Persson et al.\ 1998): HST 9177 (P182-E), with exposures of
$J$, $H$ and $K$ (5 alternate images of 15\,s for each filter).  We
also used the UKIRT Faint Standard FS30 (Casali \& Hawarden 1992;
Hawarden et al., 2001).
%
%
\section{Correction for Interstellar Extinction}
\label{DereddenSection}
To determine the intrinsic SED of the source it is necessary to
correct for interstellar extinction.  For a review of the problem see
Fitzpatrick (1999; hereafter F99).  This correction is especially
important, and problematic, in the vacuum UV.  Here the extinction is
largest, but also most uncertain, as there is significant variance
between UV extinction curves along different lines of sight; see
Fitzpatrick \& Massa (1990) for a compilation of examples.  In
correcting spectral energy distributions we would ideally like to know
the `true' extinction curve for the line of sight to the target, but
this is usually not known; instead it is common to assume some average
Galactic extinction curve, e.g.\ Seaton (1979).  A somewhat more
sophisticated approach is to select from a family of generic
extinction curves, parameterised by $R_V = A_V / E(B-V)$, based on the
properties of the line of sight (e.g.\ diffuse gas or dense cloud);
the extinction curves of Cardelli, Clayton \& Mathis (1989) and F99
adopt this approach.  If $R_V$ is not known, however, it is necessary
to assume some average value, usually taken to be $R_V = 3.1$.  As we
have no independent information on the properties of the interstellar
medium towards \target, we adopt the F99 $R_V=3.1$ extinction curve as
this should be the `best' current Galactic average curve.  In
analysing our results we must then remember that our uncertainties
depend not only on the uncertain amount of extinction but the
uncertain shape of the extinction curve.

\begin{figure}
\begin{center}
\epsfig{angle=90,width=3.4in,file=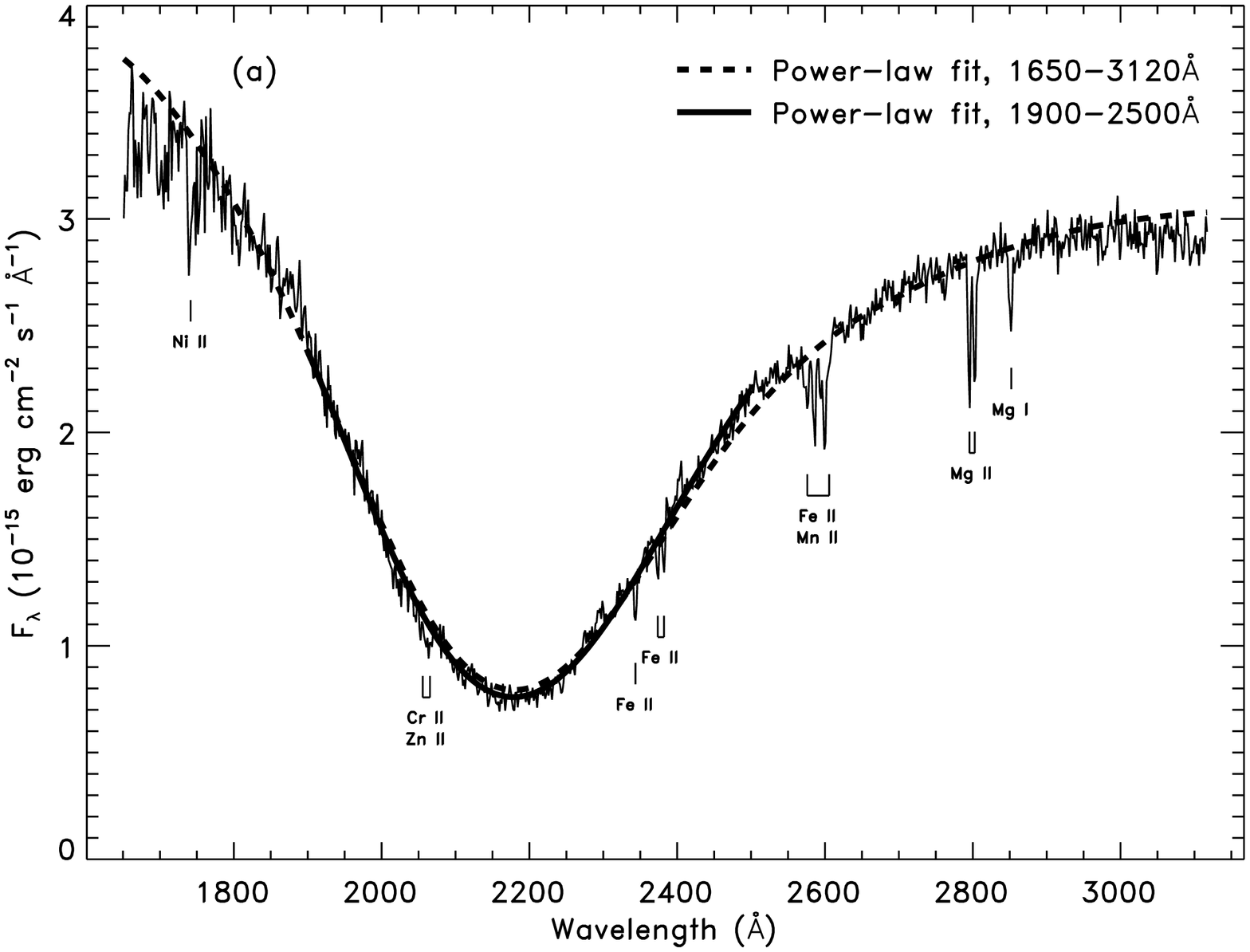}
\epsfig{angle=90,width=3.4in,file=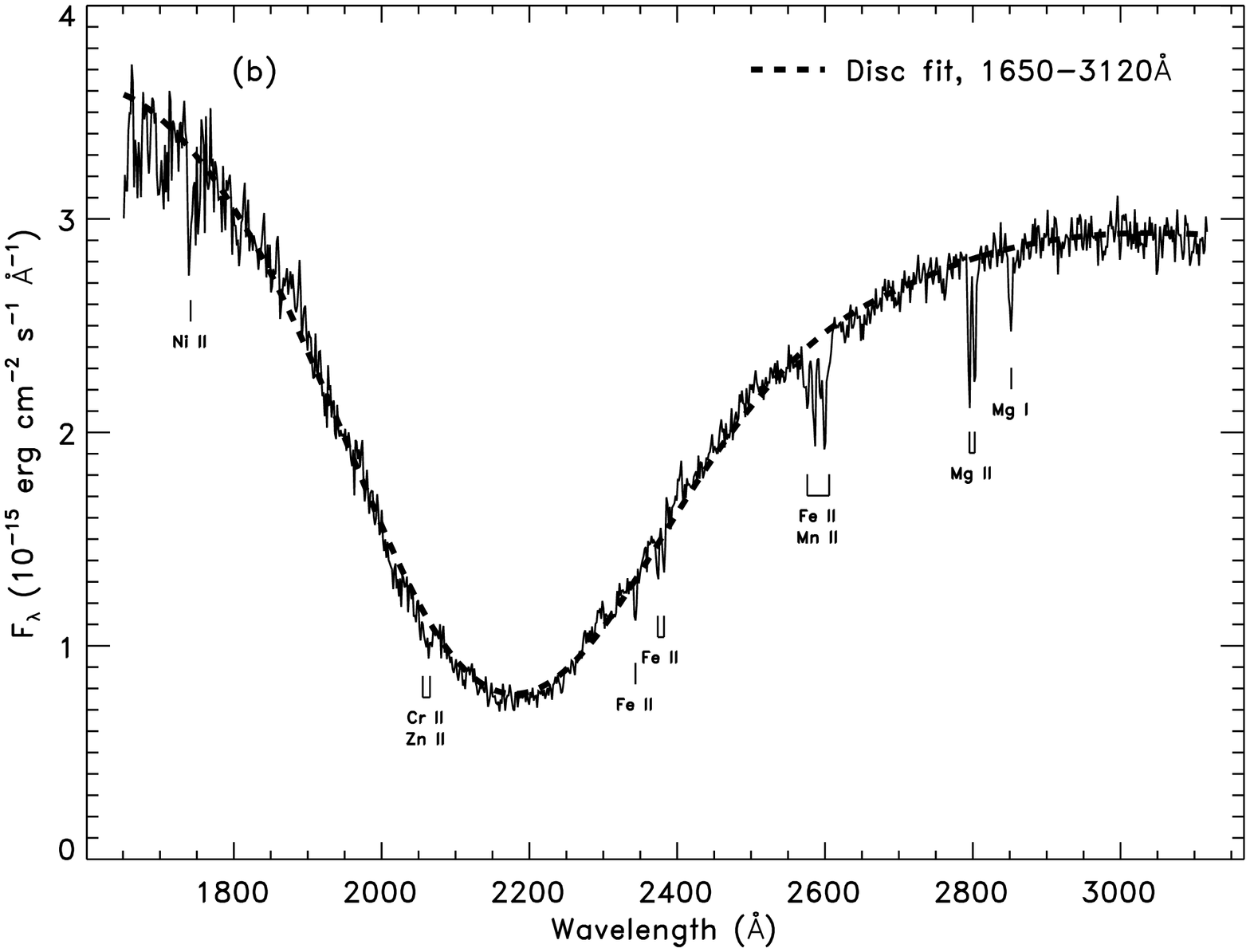}
\caption{{\bf a)} Fit to G230L spectrum using reddened power-law
models.  Two fits are shown, using wide and narrow spectral regions.
The former is clearly a poor fit.  {\bf b)} Fit using an irradiated
disc model.  This fit is good, except at the shortest wavelengths
where the extinction curve may be incorrect.  In both cases the
significant interstellar absorption lines are marked; these were
masked out before doing the fit.}
\label{DerFitFig}
\end{center}
\end{figure}

\begin{figure}
\begin{center}
\epsfig{angle=90,width=3.4in,file=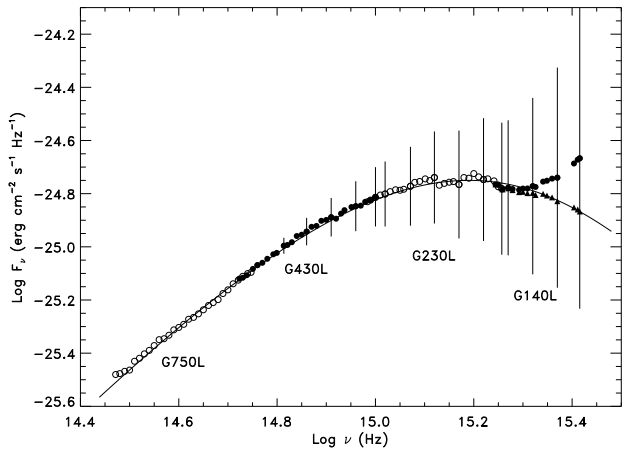}
\caption{Average spectral energy distribution from the first three
visits dereddened assuming a F99 Galactic average extinction curve
with $E(B-V)=0.58$.  Error bars on every fifth UV point indicate the
uncertainty introduced by intrinsic variance of extinction curves from
average, following the treatment in F99.  Alternate solid and filled
circles are used to distinguish the different gratings.  The solid
triangles indicate the G140L spectrum dereddened with the modified
extinction curve described in the text.  The model overplotted is an
irradiated disc model, as described in the text, with $T_{\rm out} =
18000$\,K.}
\label{GoodSpecFig}
\end{center}
\end{figure}

To estimate the amount of extinction, parameterised by $E(B-V)$ we
measure the strength of the 2175\,\AA\ interstellar absorption
feature.  This approach has previously been used on the BHXRTs X-ray
Nova Muscae 1991 (Cheng et al.\ 1992), GRO~J0422+32 (Shrader et al.\
1994), and GRO~J1655--40 (Hynes et al.\ 1998) as well as on other classes
of objects.  In fitting the feature we must assume some underlying
spectral shape, either fixed or with some free parameters, and then
adjust $E(B-V)$ to fit the data.  For \target\ we began by fitting a
reddened power-law to the data with the spectral index and
normalisation of the power-law adjusted to give the best fit for a
given $E(B-V)$.  We measure the badness-of-fit of a particular model
by the $\chi^2$ statistic of a fit to either the whole G230L spectrum
(averaged over the first three visits with spectral features masked
out) or to a subset of this.  The results of this power-law fitting
are summarised in Table~\ref{EBVTable} and two examples are plotted in
Fig.~\ref{DerFitFig}a.  Clearly fitting the whole spectral range
yielded a rather poor fit with $\chi^2_R$, the $\chi^2$ per degree of
freedom, quite high.  Using a more restricted range over which the
spectral shape will be dominated by the profile of the 2175\,\AA\
feature rather than the shape of the underlying spectrum gives a
better fit but a similar $E(B-V)$ value.  For all the fits considered
here the statistical errors in $E(B-V)$ are tiny,
$\sigma_{E(B-V)}\la0.01$, so the dominant uncertainties are
systematic.  Dereddening the SED with $E(B-V)=0.58$
(Fig.~\ref{GoodSpecFig}) it is clear why the fit to the whole G230L
spectrum is poor; the underlying spectrum does not appear to be a
power-law, but instead is curved in $\log\nu-\log F_{\nu}$ space.  As
discussed in Section~\ref{ModelSection}, a simple irradiated disc
spectrum provides a good empirical description of the SEDs from the
early visits.  We therefore repeat the fitting using such a model for
the underlying spectrum, allowing the normalisation and edge
temperature to vary to give the best fit for each $E(B-V)$.  The fits,
also summarised in Table~\ref{EBVTable} and illustrated in
Fig.~\ref{DerFitFig}b, are better than with a power-law, even when the
whole G230L spectrum is used, but the derived $E(B-V)$ values are
similar.  Hence the derived reddening is relatively insensitive to the
both the assumed spectral model and the exact wavelength range fitted.
The dominant uncertainty is due to the variation in strength of the
2175\,\AA\ feature, relative to $E(B-V)$, between different lines of
sight, about 20\,percent (F99).  Our preferred value is based on
fitting the whole G230L spectrum with an irradiated disc model.  This
provides a reasonable fit and the value derived, 0.58, conveniently
lies fairly central within the spread of inferred values, 0.56--0.60.
Our estimate of the reddening, accounting for possible variation of
the strength of the 2175\,\AA\ feature from average is then
$E(B-V)=0.58\pm0.12$.

\begin{table}
\caption{Summary of results of fitting G230L spectra with reddened models.}
\label{EBVTable}
\begin{tabular}{lccc}
\hline
Model & Fit range (\AA)  & $E(B-V)$ & $\chi^2_{\rm R}$ \\ 
\noalign{\smallskip}
Power-law & 1650--3120 & 0.56  & 1.74             \\
          & 1650--2500 & 0.59  & 1.28             \\
          & 1900--3120 & 0.57  & 1.70             \\
          & 1900--2500 & 0.59  & 1.21             \\
\noalign{\smallskip}
Irradiated disc & 1650--3120 & 0.58  & 1.24             \\
                & 1650--2500 & 0.60  & 1.18             \\
                & 1900--3120 & 0.58  & 1.26             \\
                & 1900--2500 & 0.60  & 1.20             \\
\hline
\end{tabular}
\end{table}

We show in Fig.~\ref{GoodSpecFig} the average spectrum from the first
three visits.  This has been dereddened assuming the F99 extintion
curve with $E(B-V)=0.58$.  We overplot a pure irradiated disc model
(see Section~\ref{ModelSection}) for comparison, with the irradiation
temperature inferred from the fit to the G230L spectrum.  Clearly the
model provides an extremely good fit to most of the SED, even in the
optical region which was not explicitly fitted.  The only major
deviation is in the far-UV, where the abrupt upturn is difficult to
explain.  This is in the place where a deviation would be expected if
the far-UV rise component of the extinction is somewhat different to
the Galactic average; this can vary independently of other parts of
the extinction curve, and is only weakly correlated with the strength
of the 2175\,\AA\ feature (Fitzpatrick \& Massa 1988; F99).  The
far-UV rise is also the component which differs most between different
estimates of the Galactic average extinction curve (F99).  To test
this we modify the F99 average curve by changing the far-UV rise
parameter, $c_4$.  We find that if we use $c_4=0.25$ instead of the
average value of $c_4=0.41$ then the far-UV spectrum does fit the
model very well.  $c_4=0.25$ is well within the range observed in
individual line-of-sight extinction curves (Fitzpatrick \& Massa
1990), so this is a plausible interpretation.  We also note that if
the SED is dereddened using the Seaton (1979) extinction curve, for
the same value of $E(B-V)$, then the fit to the irradiated disc model
in the far-UV is rather good (e.g.\ Hynes \& Haswell 2001).  The
deviation from the average F99 curve is thus not only within the
spread observed in single stars, but within the range of independent
estimates of the true Galactic average curve.  We therefore believe
that the most likely reason for the poor fit in the far-UV is that the
extinction curve is not quite that of F99.  For modelling in
Section~\ref{ModelSection} we therefore perform two sets of fits.  Our
preferred approach is to fit the whole dataset using the F99 average
curve except for taking $c_4=0.25$.  To test the sensitivity to the
far-UV extinction uncertainty we also fit using an unmodified F99
curve and exclude the G140L data.

%
%
\section{Spectral Modelling}
\label{ModelSection}
\subsection{The model}
To fit the broad band SED we use a very simple parameterised model.
This is based on a combination of the classic viscously heated black
body disc spectrum (Shakura \& Sunyaev 1973, Frank, King \& Raine
1992) and the modified temperature distribution for an irradiated
disc (Cunningham 1976; Vrtilek et al.\ 1990).  See these papers for
derivations of the relevent temperature distributions, and Dubus et
al.\ (1999) for a critique of the assumptions.

The model spectrum is calculated by summing a series of black bodies
over radius.  The local effective temperature of a disc annulus is
determined by the emergent flux at that radius, such that $T^4_{\rm
eff}\propto F_{\rm bol}$.  The emergent flux is the sum of viscous
energy release within that annulus, $F_{\rm bol} \propto T_{\rm
visc}^4$, and the X-rays reprocessed by the annulus, $F_{\rm
irr} \propto T_{\rm irr}^4$.  Hence the effective temperature is
\begin{equation}
T^4_{\rm eff}(R) = T^4_{\rm visc}(R) + T^4_{\rm irr}(R).
\end{equation}
For fits to the UVOIR SED the model is particularly simple, as
emission from the inner disc will not make a significant contribution;
hence the inner disc radius is not important.  The model is then
parameterised by three values, the viscous and irradiation
temperatures at the disc outer edge, $T_{\rm visc, out}$ and $T_{\rm
irr, out}$, which determine the shape of the spectrum, and the
normalisation which determines the overall flux level.  We assume that
the temperatures vary as
\begin{equation}
T_{\rm visc}(R) = T_{\rm visc, out} \left(\frac{R}{R_{\rm
out}}\right)^{-3/4}
\end{equation}
and
\begin{equation}
T_{\rm irr}(R) = T_{\rm irr, out} \left(\frac{R}{R_{\rm
out}}\right)^{-3/7}
\end{equation}
following the references given above.  After summing the local
contributions from each disc annulus an overall normalisation is
applied to match the observed fluxes.  The normalisation used is
defined as:
\begin{equation}
{\rm Norm} = \left(\frac{R_{\rm disc}}{2\times10^{11}{\rm cm}}\right)^2
\left(\frac{5{\rm kpc}}{d}\right)^2\cos i.
\label{NormDefEqn}
\end{equation}
This normalisation can inform us of {\em changes} in the outer disc
radius, although measuring an absolute radius depends on the uncertain
distance and inclination.

In principle stellar spectra could instead be used,
but the smoothness of our derived SED (e.g.\ the absence of a Balmer
jump) favours a blackbody spectrum, possibly indicating a close to
isothermal disc atmosphere (even though the disc is not expected to be
isothermal at larger optical depths).

This is obviously a very simple model, and as a theoretical model it
has shortcomings.  We will discuss the limitations in
Section~\ref{DiscussionSection}.  In spite of these objections,
variations on this model have been widely used (e.g.\ Shakura \&
Sunyaev 1973; Cunningham 1976; Vrtilek et al., 1990,1991; Cheng et
al.\ 1992; de Jong et al.\ 1996; van Paradijs 1996; King, Kolb \&
Burderi 1996; and others) and it has the advantage of few free
parameters.  We feel that this simple description of the data is more
valuable at this point than a more sophisticated model with more free
parameters, especially as this provides a good fit to the
observations.  Conclusions can be drawn about the spectral evolution,
although we must be somewhat cautious in interpreting the absolute
parameters derived.
\subsection{Model fits}
\begin{figure}
\begin{center}
\epsfig{angle=90,width=3.4in,file=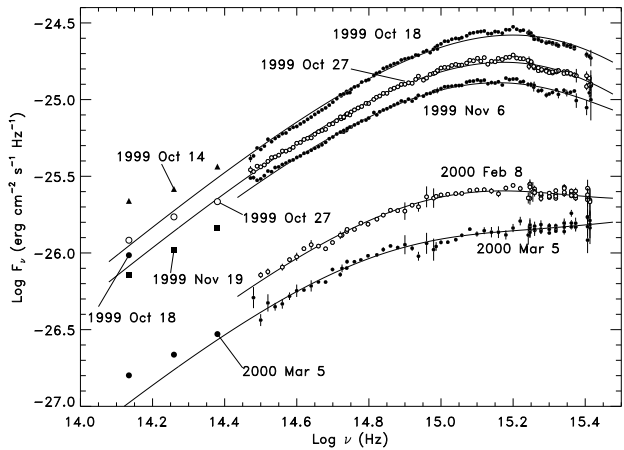}
\caption{Evolution of the UVOIR SED through the outburst.  Alternate
\HST/UKIRT visits are shown with open and filled circles for clarity.
Additional UKIRT only visits are shown with triangles or squares.  The
best fitting accretion disc model spectrum (fitted to the \HST\ data
only) is plotted on each one.  The models are only extended into the
IR for visits where we have UKIRT data.  See text for details.}
\label{SEDEvolFig}
\end{center}
\end{figure}

The broad band SEDs for all visits are shown in Fig.~\ref{SEDEvolFig}.
We have fitted the disc model described above to each.  We use only
the \HST\ data for the fit as the relative normalisation of the not
quite simultaneous UKIRT data is hard to establish without wavelength
overlap.  For each visit we have checked the overlaps between spectra
and renormalised where necessary to ensure alignment.  We also tried
using the normalisation of each spectrum as a free parameter of the
fit, but we found this introduced too much freedom and fits were being
produced with discontinuities at the boundaries between gratings.  A
fit with the normalisations fixed from the overlaps is somewhat
poorer, but less prone to spurious parameters.

All fits were done by performing a grid search in the two temperatures
and the normalisation in order to estimate confidence regions.  The
formal $\chi^2$ values for the fits are poor because the quality of
the fit is restricted by limitations in the model and extinction curve
used, rather than by statistical errors in the data.  We therefore
added an additional wavelength independent systematic error to each
point to give a best fitting $\chi^2_{\rm R}$ of 1.  These additional
errors were represented as a fixed percentage (3--5\,percent) of the
flux.  The results of the fits (with $T_{\rm visc,out}$, $T_{\rm
irr,out}$ and the normalisation all allowed to vary freely) are
summarised in Table~\ref{ModelTable} and the fits are plotted in
Fig.~\ref{SEDEvolFig}.  The confidence intervals quoted are
projections of the 3-parameter, $1\sigma$ confidence regions (Lampton,
Margon \& Bowyer 1976) estimated using the modified error prescription
described above.

\begin{table*}
\caption{Summary of results of fitting broad band SEDs with accretion
disc models.  $1\sigma$ confidence intervals are given in brackets.
These neglect uncertainties in the dereddening.  The normalisation in
the last column in defined in Eqn.~\ref{NormDefEqn}.}
\label{ModelTable}
\begin{tabular}{llrlrlrl}
\hline
Fit range (\AA) &
Date & \multicolumn{2}{c}{$T_{\rm irr,out}$ (K)} & \multicolumn{2}{c}{$T_{\rm visc,out}$ (K)} & \multicolumn{2}{c}{Normalisation} \\ 
\noalign{\smallskip}
1150--10200 &
  1999 Oct 18 & 18500 & (18300--18600) &  100 &    (0--3600) & 0.180 & (0.176--0.185) \\
& 1999 Oct 27 & 17500 & (17400--17600) &  200 &    (0--3600) & 0.142 & (0.139--0.145) \\
& 1999 Nov 6  & 15900 & (15500--16300) & 5900 & (4600--6700) & 0.130 & (0.126--0.135) \\
& 2000 Feb 8  & 12700 & (12100--13600) & 8000 & (7800--8100) & 0.035 & (0.032--0.038) \\
& 2000 Mar 5  &  8300 &  (7300--9300)  & 8000 & (7700--8200) & 0.031 & (0.027--0.035) \\
\noalign{\smallskip}
1660--10200&
  1999 Oct 18 & 20000 & (19600--20200) &    0 &    (0--6000) & 0.152 & (0.149--0.157) \\
& 1999 Oct 27 & 18100 & (16900--18500) & 5200 &    (0--8400) & 0.133 & (0.129--0.140) \\
& 1999 Nov 6  & 14700 & (13600--15800) & 8200 & (6600--9000) & 0.136 & (0.129--0.143) \\
& 2000 Feb 8  & 11400 &  (8700--13700) & 8800 & (7500--9300) & 0.037 & (0.032--0.044) \\
& 2000 Mar 5  &  7800 &     (0--10400) & 8100 & (7700--8400) & 0.031 & (0.025--0.040) \\
\hline
\end{tabular}
\end{table*}

In general the fits are rather good, surprisingly so given the
simplicity of the model.  The main difficulties are in the IR, where
the UKIRT data favour a flatter spectrum than the disc models provide,
and in the UV, where the derived SEDs are sensitive to the exact shape
of the extinction curve.  The IR problem is also hinted at in some of
the \HST\ data, for example the red end of the 1999 November 6 SED
rises somewhat with respect to the model.  We will discuss this
question further in Section~\ref{SynchrotronSection}.
\subsection{Extension to X-ray energies}
\begin{figure}
\begin{center}
\epsfig{angle=90,width=3.4in,file=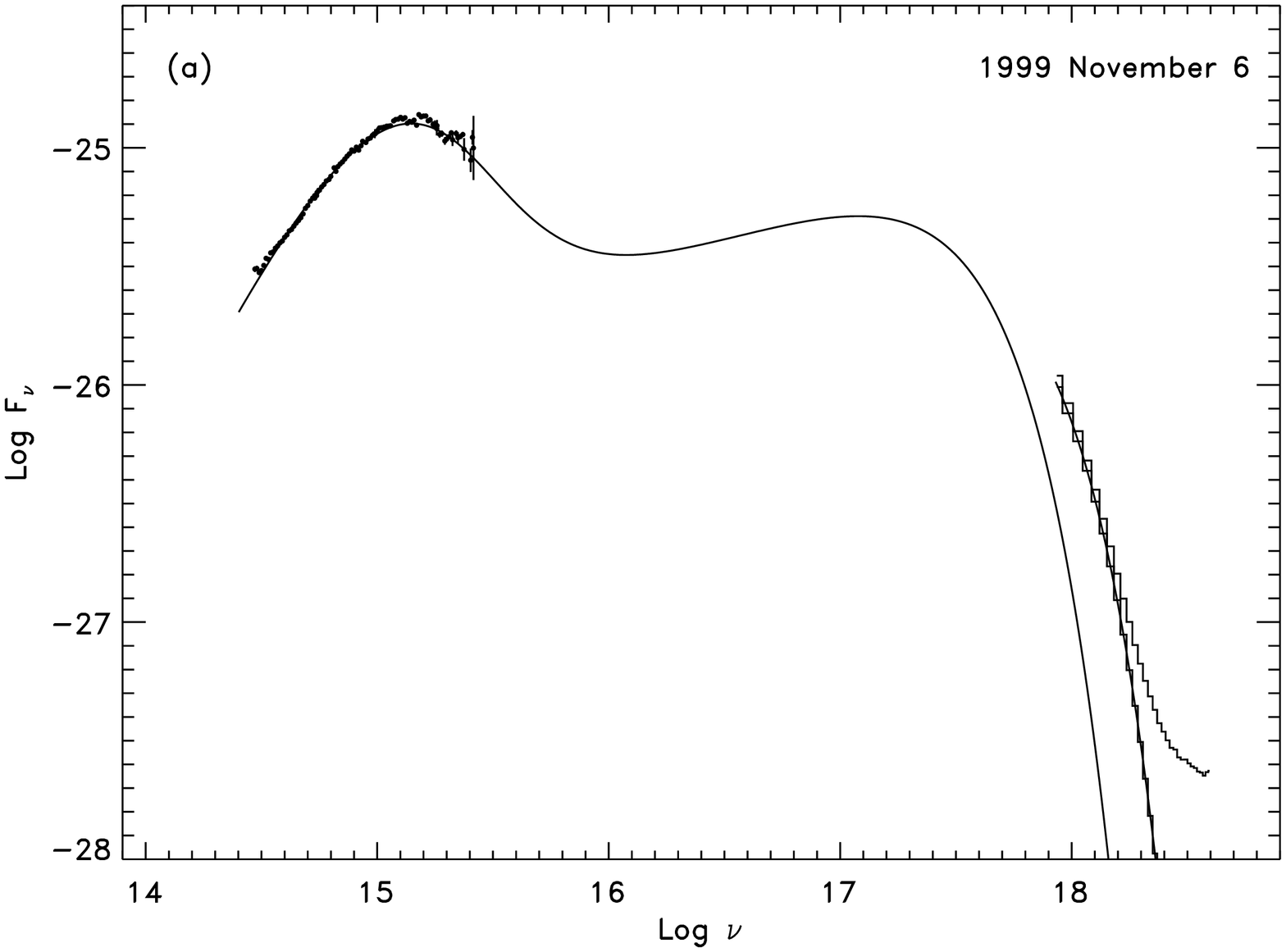}
\epsfig{angle=90,width=3.4in,file=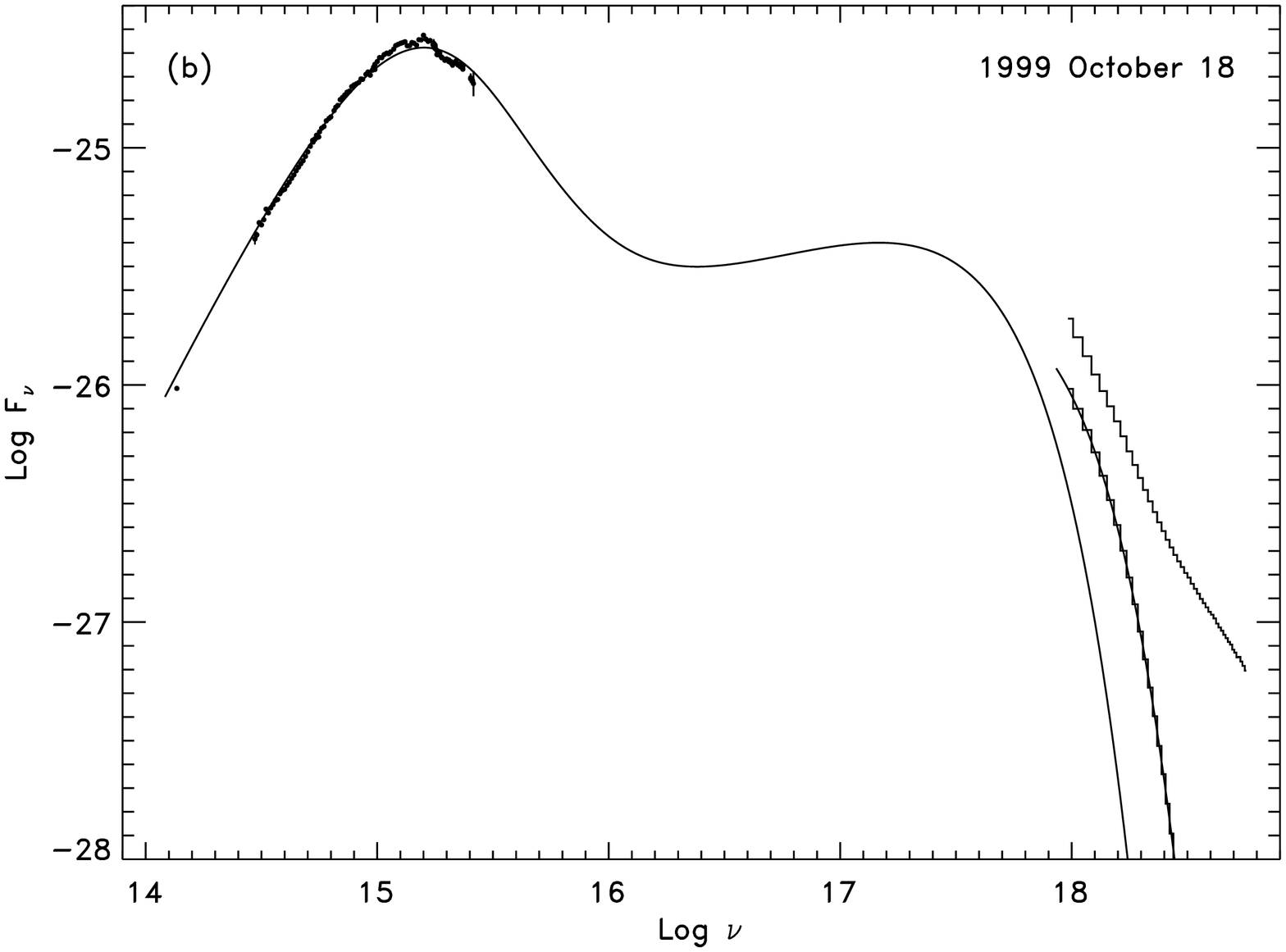}
\caption{Full X-ray--IR SED for the two visits where an X-ray disc
component was detected.  \HST\ and UKIRT points are indicated at
left.  Two \RXTE\ spectra are given in each case; the upper is the
total spectrum, the lower is the decomposed disc component.  The
solid line is the disc model SED; the short additional segment in the
X-rays is the same modified by spectral hardening.}
\label{FullSEDFig}
\end{center}
\end{figure}

A natural progression from fitting the UVOIR SEDs is to also require
that the disc models be consistent with the X-ray disc component,
if seen.  Our \RXTE\ data do show disc components for the early
epochs, 1999 October 18 and November 6.  This component dominates in
the latter visit.  As the viscous temperature is also rather poorly
constrained by the UVOIR data from the earlier visit, the November 6
data should provide the strongest test of consistency.  To extend the
model to X-rays we require an additional parameter, $r_{\rm in}$.  In
practice, given the uncertain distance and system parameters the data
do not strictly constrain this, but rather $r_{\rm in} / r_{\rm out}$.
Adjusting this ratio produces a one-parameter family of curves, with
varying cutoff energies, one of which should match the X-ray disc
component, assuming that a common model is applicable.  There is some
freedom to adjust the normalisation of these curves by varying $T_{\rm
visc}$ within the uncertainties allowed by the UVOIR fits.  We find
that this does appear to work, provided that the X-ray part of the SED
is modified by spectral hardening (Shimura \& Takahara 1995).  We use
a hardening factor of $f_{\rm col}=1.7$ for both epochs, although this
is a crude approximation as discussed below.  Fig.~\ref{FullSEDFig}a
shows the X-ray and UVOIR data from November 6 with the best fitting
model.  This model has $T_{\rm visc}=5040$\,K (consistent with the
UVOIR fits), $r_{\rm in} / r_{\rm out}=7.2\times10^{-5}$ and other
parameters as given in Table.~\ref{ModelTable}.  The bolometric disk
luminosity of this model is $L_{\rm bol} \sim
1.6\times10^{38}$\,erg\,s$^{-1}$ for a distance $\sim7.6$\,kpc (see
Section~\ref{DistanceSection}), corresponding to $\sim10$\,percent of
$L_{\rm Edd}$ for a 10\,M$_{\odot}$ black hole.  There will be a small
additional contribution from the power-law component.

The same exercise can be carried out for the October 18 data, although
here, the UVOIR constraints on $T_{\rm visc}$ are much weaker so there
is more freedom to adjust this to fit the X-ray data.  The best common
fit is shown in Fig.~\ref{FullSEDFig}b, corresponding to $T_{\rm
visc}=3950$\,K and $r_{\rm in} / r_{\rm out}=3.9\times10^{-5}$.  The low
value of $T_{\rm visc}$ derived is almost consistent with the
$1\sigma$ limit inferred from the UVOIR SED.  The bolometric disc
luminosity of this model is also $L_{\rm bol} \sim
1.6\times10^{38}$\,erg\,s$^{-1}$, although for this visit the
power-law component will make a much larger contribution, so the total
luminosity is higher than for November 6.

Assuming a disc radius $\sim0.9 \times 2\times10^{11}$\,cm, consistent
with likely parameters and a tidally truncated disc, then the implied
inner radius at the first visit is $\sim90$\,km.  This is about
$3R_{\rm Sch}$ for a 10\,M$_{\odot}$ black hole, suggesting a disc
extending to approximately the last stable orbit.  Without a more
reliable distance or system parameters, it is obviously not worth
trying to be more precise.  The derived inner radius is also sensitive
to the choice of spectral hardening factor $f_{\rm col}$, and contrary
to what is often assumed, this is unlikely to be constant; Merloni et
al.\ (2000) suggest a range of 1.7--3 is plausible.  The implied inner
radius for the 1999 November 6 visit is somewhat larger than for
October 18.  Given the dramatic change in the spectral state,
however, we cannot confidently say that this change is real (c.f.\
Merloni et al.\ 2000); it could reflect a change in $f_{\rm col}$.

We therefore conclude that on those visits where X-ray disc
emission is measured, it is consistent with a plausible extrapolation
of the UVOIR fits, assuming a disc extending to around the last stable
orbit.  Therefore the disc models considered, modified for spectral
hardening where necessary, and with additional non-thermal hard X-ray
and IR contributions, represent an acceptable fit to the whole
X-ray--IR SED.

\subsection{The distance to the source}
\label{DistanceSection}
The normalisation of the disc fits is dependent on the distance to the
source, although obviously other factors also affect it.  The
normalisation we used is defined by Eqn.~\ref{NormDefEqn}.  To
determine the distance we need to know the binary parameters.  What we
actually know is $P_{\rm orb}=(9.16\pm0.08)$\,hr and
$f(M)=(7.4\pm1.1)$\,M$_{\odot}$ (Filippenko \& Chornock 2001).  No
eclipses have been reported, although an 0.4\,mag ellipsoidal
modulation appears to be present (Sanchez-Fernandez et al.\ 2000).
The inclination is thus likely to be moderately large.  Filippenko \&
Chornock (2001) found a best fit spectral type of G5, although they
did not use earlier type templates.  We can make a plausible distance
estimate by assuming a typical black hole mass of
$M_1\sim10$\,M$_{\odot}$, a G5 companion mass of
$M_2\sim1$\,M$_{\odot}$ (Gray 1992), $P_{\rm orb}=9.16$\,hrs,
$i\sim60^{\circ}$ and a disc filling 90\,percent of the black hole's
Roche lobe (i.e.\ tidally truncated) for the first visit with the
largest normalisation.  We then derive a distance estimate of
7.6\,kpc.  Obviously the errors on this are large.  To attempt to
quantify these we perform a simple Monte Carlo simulation.  We
construct a population of binaries with randomly chosen parameters.
We take a uniform black hole mass distribution of
$M_1=5-12$\,M$_{\odot}$ (Bailyn et al.\ 1998), a uniform companion
mass distribution ($M_2=0.68-1.12$\,M$_{\odot}$) corresponding to
G0--G9 spectral type (Gray 1992), allowing for a possibly undermassive
companion (Kolb, King \& Baraffe 2001), a Gaussian period distribution
of $P_{\rm orb}=(9.16\pm0.08)$\,hr and a uniform distribution in $\cos
i=0-1$.  We then reject any binaries in which the central source is
eclipsed and weight the remainder so as to produce a Gaussian
mass-function distribution of $f(M)=(7.4\pm1.1)$\,M$_{\odot}$.  The
resulting distribution of distance estimates then has a 95\,percent
confidence range of 4.6--8.0\,kpc, so we believe this is a reasonable
estimate of the range of probable distances given the current
uncertainty in the system parameters.  Of course these estimates are
model dependent, and we have not accounted for further uncertainties
introduced by assuming an extinction curve and $E(B-V)$ value, so a
closer or further distance cannot confidently be ruled out.  Also if
the disc does not extend to the tidal truncation radius ($0.9r_{\rm
lobe}$) at maximum then the distance would be reduced.

\subsection{Physical parameters}
\label{ParameterSection}

We can now attempt to use the assumed binary parameters and our
distance estimate to convert the viscous temperatures and
normalisations derived into mass transfer rates, although there
obviously will be large uncertainties in this process.  The mass
transfer rate, $\dot{M}$, is related to $T_{\rm visc}$ and the
normalisation by (see Frank et al.\ 1992 and Eqn.~\ref{NormDefEqn})
\begin{eqnarray}
\dot{M} & = & 4.54\times10^{-7}
\left(\frac{M}{10M_{\odot}}\right)^{-1}
\left(\frac{T_{\rm visc}}{10^4\,{\rm K}}\right)^4
\left(\frac{d}{10\,{\rm kpc}}\right)^3\nonumber\\ & &
\times \left(\frac{{\rm Norm}}{\cos i}\right)^{-3/2}
{\rm M_{\odot}}\,{\rm yr^{-1}}
\end{eqnarray}
For the first two visits $T_{\rm visc}$ is not well defined, but
$\dot{M}$ can be estimated for the other visits.  Assuming
$M=10$\,M$_{\odot}$, $d=7.6$\,kpc and $i=60^{\circ}$, we derive
$\dot{M}\sim2.6\times10^{-8}$, $1.2\times10^{-8}$ and
$1.0\times10^{-8}$\,M$_{\odot}$\,yr$^{-1}$ for 1999 November 6, 2000
February 8 and 2000 March 5 respectively.  Assuming a 10\,M$_{\odot}$
black hole and an accretion efficiency of 10\,percent, the Eddington
limited mass transfer rate is $\dot{M}_{\rm
Edd}=2.3\times10^{-7}$\,M$_{\odot}$\,yr$^{-1}$.  Thus the November 6
observation corresponds to $\dot{M}\sim0.1\dot{M}_{\rm Edd}$, as
already estimated from the X-ray spectrum, while the later visits are
lower, $\sim5$\,percent.  These are reasonable numbers.  The X-ray
decay in this period is more dramatic than a factor of 2--3, but this
will be amplified by the shifting of the X-ray spectrum out of the ASM
bandpass.

For the 1999 November 6 visit it is also of interest to compare the
irradiation temperature with the X-ray luminosity.  As will be
discussed in Section~\ref{LimitationSection}, some of the assumptions
used deriving the exact irradiation temperature distribution are
unjustified, and so it would not be appropriate to interpret the
results in too much detail.  A simple prescription has been advanced
by Dubus et al.\ (1999), who write the irradiation temperature as
\begin{equation}
T_{\rm irr}^4 = C \frac{\dot{M}c^2}{4\pi \sigma R^2}
\end{equation}
where a simple relation between $\dot{M}$ and $L_{\rm X}$ has been
assumed.  The parameter $C$ encompasses our ignorance of the exact
irradiation geometry; Dubus et al.\ (1999) use $C\sim5\times 10^{-4}$
for consistency with earlier results.  Using our parameters derived
for 2000 November 6 we obtain $C\sim7.4\times10^{-4}$, in reasonable
agreement with the value used by Dubus et al.\ (1999).  This agreement
is as good as can be expected given the uncertainty about many
parameters.
%
%
\section{Discussion}
\label{DiscussionSection}

\subsection{Interpretation in the context of the disc instability model}

The decrease in the normalisations given in Table~\ref{ModelTable}
clearly suggest a systematic decrease in the projected disc area as
the outburst decays (Eqn.~\ref{NormDefEqn}); as the distance and
inclination are constant it must be the projected area which varies.
We attribute this to changes in radius of the hot area of the disc,
but changes in the projection due to disc warping may also be involved
(c.f.\ Section~\ref{WarpSection}).  This could represent a change in the
actual radius of the outer disc, but given the relatively large change
seen it is more likely that it represents a cooling wave moving
inwards through the disc.  The spectrum would then be dominated by the
inner hot disc; the outer disc is present but is much cooler and
possibly optically thin, so would only make a weak contribution.  King
\& Ritter (1998) have considered the evolution of an irradiated disc
in a BHXRT.  Their model interprets two kinds of decay behaviour --
exponential and linear decays.  During the exponential decay phase the
whole disc is maintained in a hot state by X-ray irradiation and the
exponential decay is due to the draining of mass from the disc.
During the linear decay phase a cooling wave moves inwards at a rate
determined by the decreasing irradiation flux.  In a typical
short-period BHXRT such as \target, an exponential decay phase is
followed by a linear decay; all of our observations took place during
the exponential decay phase (Casares et al.\ 2000 report photometry
indicating that the exponential decay phase lasted until 2000 May 30).
Nonetheless we see a significant decrease in hot disk area suggesting
that actually the disc is not all maintained in the hot state, but
that the cooling wave is allowed to propagate.  Further, we see the
irradiation temperature at the edge of the disc decreasing, suggesting
that the cooling wave propagation is not controlled straightforwardly
by the decreasing irradiation flux -- neither the irradiation
temperature, nor the combined effective temperature, is approximately
constant at the edge of the hot disc.  Our observations thus do not
support the details of the model of King \& Ritter (1998) in which a
cooling front is completely inhibited.

If the decrease in normalisation is due to the propagation of a
cooling wave then the implied velocity is rather slow;
$\sim1.5\times10^{11}$\,cm over $\sim140$\,days, corresponding to an
average of only $\sim6$\,cm\,s$^{-1}$.  This is significantly below
theoretical cooling wave velocities (in the absence of irradiation) of
$\sim1$\,km\,s$^{-1}$ (Menou, Hameury \& Stehle 1999; Cannizzo 2001).
Given the present of a secondary maximum (around day 140 in
Fig.~\ref{ASMFig}) which may involve restarting of the cooling wave
such a simple calculation is naive.  Nonetheless, a discrepancy with
theory is hard to avoid.  It therefore seems likely that irradiation
does slow the cooling wave significantly; in this respect we support
King \& Ritter (1998).  The observations, however, suggest a
modification of the model in which irradiation does not completely
stop the cooling wave, but allows it to move slowly as the irradiating
flux declines.

As noted earlier, given the uncertain extinction curve and simplistic
local spectrum assumed, we should not interpret the temperatures
described too literally.  Nonetheless it is worth noting that they do
qualitatively make sense in the context of the disc instability model.
In the later visits when irradiation is less strong, the inferred
viscous temperatures at the edge of the hot disc, $\sim8000$\,K are
close to that expected; the viscous temperature is the best indicator
of the internal temperature of the disc, since irradiation will only
ever be dominant in the surface layers (c.f.\ Dubus et al.\ 1999).
For $\alpha \sim 0.1$, $M_1\sim10$\,M$_{\odot}$ and
$R\sim8\times10^{10}$\,cm (assuming a first visit disc radius
$\sim0.9\times2\times10^{11}$\,cm and normalisation $\propto R^2$) we
expect a critical viscous effective temperature of $\sim6600$\,K
(Dubus et al.\ 1999).  For the first two visits, however, the inferred
viscous temperature is lower than this, and the weak dependence of the
critical temperature on radius cannot resolve this discrepancy.  It
seems likely that the strong irradiation inferred at these epochs is
stabilising the disc at a radius larger than would be expected from
viscous heating alone; as discussed by Dubus et al.\ (1999)
strong irradiation is expected to reduce the critical temperature.

The evolution of the spectrum from irradiation dominated to viscous
heating dominated is a consequence of the weaker dependence of
irradiation on radius ($\propto R^{-2}$) than the dependence of
viscous heating ($\propto R^{-3}$).  For a given irradiating flux and
a large enough disc the outer part will be irradiation dominated
whereas the inner part is viscous heating dominated.  Analogously, we
would expect the edge of a large disc to be irradiation dominated
whereas the edge of a small disc will be dominated by viscous heating,
assuming that the ratio of irradiating flux to $\dot{m}$ does not
change dramatically.

\subsection{Disc warping}
\label{WarpSection}

A further factor that may come into play is warping of the disc.
Dubus et al.\ (1999) have already invoked this as a possible mechanism
to allow irradiation of the outer parts of a disc when it would
otherwise be self-shielded.  If warping is important then the
irradiated area (and hence the normalisation of the SED) may depend in
a complex way on the warp geometry and changes in the normalisation
could indicate changes in this geometry rather than in the disc, or
cooling front, radius.  Such changes have previously been suggested to
explain the lightcurves of another BHXRT with more unusual behaviour,
GRO~J1655--40 (Esin, Lasota \& Hynes 2000).  Ogilvie \& Dubus (2001)
have argued, however, that short-period BHXRTs with `classic'
fast-rise, exponential-decay (FRED) lightcurves, such as \target, are
stable against radiation driven warping and that it will only be
important for longer period systems (e.g.\ GRO~J1655--40).  Clearly
the effect of a warp, if present, would be complex, and the observed
spectrum may depend on the precession phase at the time of each
observation.  Also for a transient in outburst the warp is likely to
be evolving as the outburst progresses.  So although the development
of a warp is unlikely to play a large role in a system such as
\target, we cannot observationally rule out this possibility that
changes in warp geometry could contribute to the spectral changes we
see.

\subsection{Limitations of the model}
\label{LimitationSection}

Using blackbody local spectra is obviously an extreme simplification.
It does work rather well, however, and the observed SEDs are almost as
smooth as black bodies, in particular there is no detectable Balmer
jump.  This means that the obvious next refinement, to use stellar
atmosphere spectra, as done by Vrtilek et al.\ (1990, 1991), would
give a worse fit, since stars in the critical 10,000--20,000\,K range,
corresponding to the temperatures dominant in the disc spectrum, have
strong Balmer absorption.  A more sophisticated treatment would,
therefore, require use of a model atmosphere generated specifically
for an irradiated accretion disc atmosphere.  Development of such
models is underway (Hubeny 2001), but they are still in their infancy,
depend on much uncertain physics and introduce many more free
parameters.  Since a blackbody model does provide an acceptable fit
to the data we feel this is the most appropriate choice at this time.

There is also little justification for the irradiation temperature
distribution assumed.  The derivation of $T_{\rm irr} \propto
R^{-3/7}$ assumes a vertically isothermal disc and cannot be correct.
King, Kolb \& Szuszkiewicz (1997) argue for a more realistic solution
with the vertical temperature gradient accounted for.  Their model
gives $T_{\rm irr} \propto R^{-0.40}$, compared to $R^{-0.43}$ which
we have used.  The self-consistent treatment of Dubus et al.\ (1999)
actually predicts that the outer disc should not be irradiated at all,
contrary to observations (e.g.\ van Paradijs \& McClintock 1995).
Consequently additional factors such as disc warping and/or scattering
of X-rays must be important and the true dependence of $T_{\rm irr}$
on radius is not known.  Dubus et al.\ (1999) write the irradiation
law as $T_{\rm irr} \propto C R^{-1/2}$, where the explicit radial
dependence comes only from the inverse square law and additional
unknown geometric factors are included in $C$, which they assume to be
constant.  In practice, however, the difference between assuming
$R^{-0.5}$, $R^{-0.40}$ and $R^{-0.43}$ dependences is rather small.
We have tested using all of these and our conclusions are not
significantly affected by which we choose.  The $R^{-3/7}$ is the
intermediate relation and the most widely used to date, so is a
reasonable approximation to choose.

Another simplification made is that a fixed fraction of incident
X-rays are reprocessed into UVOIR emission, i.e.\ that the albedo is
constant.  This will not be true in the inner disc where the
opacity may become dominated by electron scattering and a larger
fraction of X-rays will be Compton reflected.  In this inner region
the disc will behave more like a mirror and reprocessing will be a
less effective form of heating (A. R. King, 2001, priv.\ comm.;
A. C. Fabian, 2001, priv.\ comm.).  Shakura \& Sunyaev (1973) give an
approximate expression for the radius within which electron scattering
will dominate, $R/3R_{\rm sch} \sim 6.3\times10^3 \dot{m}^{2/3}$,
where $\dot{m}$ is the mass transfer rate in units of the critical
(Eddington limited) rate.  We have experimented with turning off
irradiation within this radius in our model.  For a typical BHXRT in
outburst with $\dot{m}\sim0.1-1.0$, the effect on the UVOIR SED is
relatively modest, $\sim10$\,percent in the far-UV dropping rapidly
with increasing wavelength.  Consequently it will not be possible to
disentangle this effect without a better extinction curve as it will
distort the spectrum in a similar way to the far-UV rise.

\subsection{The infrared spectrum}
\label{SynchrotronSection}
The infrared photometry suggests a flatter spectrum than an
extrapolation of the disc models predicts.  This is particularly
prominent on the first UKIRT observation where the flattening in the
$K$ band appears quite pronounced.  Brocksopp et al.\ (2001) have
independently suggested that flat-spectrum synchrotron emission may be
important in the IR in this source, and our results support their
conclusion.  In fact, another BHXRT, XTE~J1118+480, shows a very
convincing case for flat-spectrum synchrotron emission in the IR, and
probably also in the optical (Hynes et al.\ 2000; Chaty et al.\ in
preparation).  It is interesting that the IR spectral slope does not
change as it drops (except possibly for the first observation on the
outburst rise).  This would also fit with synchrotron, for example
with the power of a jet changing but the spectrum staying about the
same.

Of course an alternative explanation might be that the IR excess comes
from the cool outer disc which we expect to be present, at least in
the later visits.  We would, however, expect this cool disc
contribution to get stronger as the hot disc shrinks.  The presence of
the excess in all observations with $JHK$ coverage does not fit with
this, so the synchrotron interpretation seems more likely.
%
%
\section{Conclusion}
\label{ConclusionSection}
We have analysed the broad band UVOIR and X-ray spectra of \target\ in
outburst.  We have found that all of the observations can be accounted
for with a simple model accretion disc heated by internal viscosity
and X-ray irradiation, if additional non-thermal components provide
the flatter IR component and the X-ray power-law.  As the outburst
declines we see a decrease in the normalisation of the disc models and
the irradiation temperature at the edge of the hot disc and an
increase in the edge viscous temperature.  We therefore see the outer
disc change from being irradiation dominated towards being
viscous-heating dominated with an accompanying decrease in the
emitting area.  We interpret the latter as evidence for a cooling
wave; the transition from irradiative to viscous heating is
qualitatively consistent with this interpretation.  The evolution is
not consistent with a cooling wave controlled by some fixed
irradiation temperature, but appears to depend on a combination of
irradiative and viscous heating.
%
%
\section*{Acknowledgements}
RIH would like to thank Guillaume Dubus for some interesting
discussions on irradiation of accretion discs and Jim Pringle and
Andrew King for constructive criticism of these results.  RIH, CAH and
SC acknowledge support from grant F/00-180/A from the Leverhulme
Trust.  Support for proposal GO\,8245 was provided by NASA through a
grant from the Space Telescope Science Institute, which is operated by
the Association of Universities for Research in Astronomy, Inc., under
NASA contract NAS 5-26555.  Thanks to Tony Roman and Kailash Sahu at
STScI for support.  The United Kingdom Infrared Telescope is operated
by the Joint Astronomy Centre on behalf of the U.K. Particle Physics
and Astronomy Research Council.  UKIRT Service observations were
obtained thanks to override time which was pre-approved in case of
outbursting transients (U/98A/19), to be coordinated with our \HST\
and \XTE\ observations.  We thank John K. Davies who scheduled,
coordinated, prepared and supervised the override observations.  We
also thank Ian Smail and Glenn Morrison (Oct 13 observations), Rob
J. Ivison (Oct 18), A. Adamson (Oct 27) and T. R. Geballe (Nov 19) who
performed the mentioned observations and Ken Chambers for obtaining
the last visit IR coverage during IfA time.  This research has made
use of the SIMBAD database, operated at CDS, Strasbourg, France, the
NASA Astrophysics Data System Abstract Service and quick-look results
provided by the ASM/\XTE\ team.
%
%

%
\end{document}